\documentclass[a4paper,10pt]{article}
\usepackage[ams]{}
\usepackage{amssymb}
\usepackage{graphicx}
\usepackage[theorem]{}
\begin{document}
\title{The Geometry of Jordan Matrix Models}
\author{Michael Rios\footnote{email: mrios4@calstatela.edu}\\\\\emph{California State University, Los Angeles}\\\emph{Mathematics Graduate Program}
\\\emph{5151 State University Drive}\\\emph{Los Angeles, CA 90032-8531}  } \date{\today}\maketitle
\begin{abstract}
We elucidate the geometry of matrix models based on simple formally real Jordan algebras.  Such Jordan algebras give rise to a nonassociative geometry that is a generalization of Lorentzian geometry.  We emphasize constructions for the exceptional Jordan algebra and the exceptional Jordan $C^*$-algebra and describe the projective spaces related to the exceptional cubic matrix model and the $E_6$ matrix model.  The resulting projective spaces are shown to be exceptional versions of projective twistor space, thus revealing the existence of exceptional twistor string theories that are dual to octonionic matrix models.\\\\

$Keywords:$ Jordan algebras, matrix theory, twistor string theory.
\end{abstract}
\newpage
\tableofcontents
\newpage
\section{Introduction}
In recent years, various matrix models have been proposed as nonperturbative definitions for M-theory \cite{3,4,5,6}.  A common thread linking these matrix models is the use of Hermitian matrices over normed division algebras.  Such Hermitian matrices form \emph{Jordan algebras} \cite{7,19,26} when the usual matrix product is replaced by the Jordan product.  Jordan algebras are commutative and nonassociative, and this leads to a \emph{nonassociative geometry} that is related to Jordan algebras in much the same way that noncommutative geometry is related to noncommutative $C^*$-algebras.\\
\indent In this paper we review the nonassociative geometries arising from the simple formally real Jordan algebras.  The simple formally real Jordan algebras include all Hermitian matrix algebras over $\mathbb{R}$, $\mathbb{C}$, $\mathbb{H}$, and $\mathbb{O}$.  Thus our geometric analysis is sufficiently rich to be applicable to a wide class of matrix models.  We place particular emphasis on the exceptional Jordan algebra $\mathfrak{h}_3(\mathbb{O})$, which is the self-adjoint part of the exceptional Jordan $C^*$-algebra $\mathfrak{h}_3(\mathbb{C}\otimes\mathbb{O})$ \cite{20}.  We show the nonassociative geometries of $\mathfrak{h}_3(\mathbb{O})$ and $\mathfrak{h}_3(\mathbb{C}\otimes\mathbb{O})$ yield Jordan generalizations of Lorentzian geometry, with light cone geometries that are exceptional analogs of projective twistor spaces.

\section{Jordan Algebras and Projective Spaces}
In this section we review the properties of Jordan algebras and their relationship to projective spaces \cite{7, 19}.
\subsection{Jordan algebras}

\textbf{Definition 2.1.1}\\  A \textbf{Jordan algebra} $\mathcal{J}$ is a real vector space $\mathcal{J}$ equipped with the Jordan product (i.e.  a bilinear form) $(a,b)\rightarrow a\circ b$ satisfying $\forall a,b\in \mathcal{J}$:
\begin{center}$a\circ b = b\circ a$,\end{center}
\begin{center}$a\circ (b\circ a^2) = (a\circ b)\circ a^2$. \end{center}
$\mathcal{J}$ is \textit{unital} if it admits a unit with respect to the Jordan product.  Other useful operations include trace, determinant, and the \emph{Freudenthal product}:
\begin{equation}
a\ast b = a\circ b -\frac{1}{2}a\hspace{.1cm}\textrm{tr}(b) - \frac{1}{2}b\hspace{.1cm}\textrm{tr}(a) - \frac{1}{2} I(\textrm{tr}(a\circ b)- \textrm{tr}(a)\textrm{tr}(b)).
\end{equation}
\\
\textbf{Definition 2.1.2}\\
A \textbf{formally real} Jordan algebra $\mathcal{A}$ is a Jordan algebra satisfying for all $n$:
\begin{displaymath}
a^2_1+\dots+a^2_n=0\Rightarrow a_1=\dots=a_n=0.
\end{displaymath}

\indent An \textit{ideal} in a Jordan algebra $\mathcal{J}$ is a subspace $\mathcal{B}\subseteq \mathcal{J}$ such that $b\in \mathcal{B}$ implies $a\circ b\in \mathcal{B}$ for all $a\in\mathcal{J}$.  A Jordan algebra $\mathcal{J}$ is \textit{simple} if its only ideals are {0} and $\mathcal{J}$ itself.  A Jordan algebra $\mathcal{J}$ is \textit{special} if there exists a monomorphism $\sigma$ of $\mathcal{J}$ into an associative algebra $\mathcal{A}^+$ \cite{26}.  Jordan algebras that are not special are called \textit{exceptional}.\\
\indent The simple formally real Jordan algebras consist of four families and one exception:
\begin{displaymath}
1.\indent \mathfrak{h}_n(\mathbb{R})
\end{displaymath}
\begin{displaymath}
2.\indent \mathfrak{h}_n(\mathbb{C})
\end{displaymath}
\begin{displaymath}
3.\indent \mathfrak{h}_n(\mathbb{H})
\end{displaymath}
\begin{displaymath}
4.\indent \mathbb{R}^n\oplus \mathbb{R}
\end{displaymath}
\begin{displaymath}
5. \indent\mathfrak{h}_3(\mathbb{O})
\end{displaymath}
where $\mathfrak{h}_n(\mathbb{K})$ denotes the algebra of $n\times n$ Hermitian matrices with entries from $\mathbb{K} = \mathbb{R}, \mathbb{C}, \mathbb{H}$ and product $a\circ b = \frac{1}{2}(ab+ba)$.  The family $\mathbb{R}^n\oplus \mathbb{R}$ consists of Jordan algebras called \emph{spin factors} \cite{7}.  $\mathfrak{h}_3(\mathbb{O})$ is the exceptional Jordan algebra of $3\times 3$ Hermitian matrices with octonionic entries, used by Jordan, Wigner and von Neumann for a generalized quantum mechanics in 1934 \cite{1}.
\subsection{Projective Spaces}
\textbf{Definition 2.2.1}\\
A \emph{projective n-space} $\mathbb{KP}^n$ over a field $\mathbb{K}$ has points that are 1-spaces of an $(n+1)$-dimensional vector space over the field $\mathbb{K}$.  It is coordinatized by non-zero $(n+1)$-tuples $(x_0,x_1,...,x_n)$ with the understanding that they are equivalent to $(\lambda x_0, \lambda x_1, ..., \lambda x_n)$ for $\lambda\neq 0$ \cite{14}.\\\\
\indent Definition 2.2.1 suffices for $\mathbb{R}$ and $\mathbb{C}$, and generalizes well for the quaternions $\mathbb{H}$.  However, it fails for the octonions $\mathbb{O}$ \cite{6,14}.  Fortunately, there is an alternative definition of projective $n$-space \cite{14}, where points are represented by Jordan projection operators onto the corresponding 1-spaces of $\mathbb{K}^{n+1}$.\\\\
\textbf{Definition 2.2.2}\\
A \textbf{projective \textit{n}-space} $\mathbb{KP}^n$ over a normed division algebra $\mathbb{K}$ has points that are rank one projections of the simple formally real Jordan algebra $\mathfrak{h}_{n+1}(\mathbb{K})$, i.e., $p\in\mathfrak{h}_{n+1}(\mathbb{K})$ such that $p^2=p$ and $tr(p)=1$.  Lines are given by rank two projections $p\in\mathfrak{h}_{n+1}(\mathbb{K})$ that satisfy $p^2=p$ and $tr(p)=2$.\\\\
A point $p_1$ lies on a line $p_2$ just when $p_1\circ p_2=0$ \cite{14}.  By commutivity, we also have $p_2 \circ p_1 = 0$, so `lies on' is a symmetric condition.\\\\
\textbf{Example 2.2.1 ($\mathbb{CP}^3$ Twistor Theory and $\mathfrak{h}_4(\mathbb{C})$)}\\
\indent By definition 2.2.2, points of the projective twistor space $\mathbb{CP}^3$ are rank one projections of $\mathfrak{h}_4(\mathbb{C})$.  Extending Penrose's construction in \cite{39}, we recover rank one projections of $\mathfrak{h}_4(\mathbb{C})$ via the procedure:
\begin{displaymath}
p=\lambda\lambda^{\dagger}=\left(\begin{array}{c}z_1 \\ z_2 \\ z_3 \\ z_4 \end{array}\right)\left(\begin{array}{cccc}\overline{z}_1 & \overline{z}_2 & \overline{z}_3 & \overline{z}_4 \end{array}\right) = \left(\begin{array}{cccc}z_1\overline{z}_1 & z_1\overline{z}_2 & z_1\overline{z}_3 & z_1\overline{z}_4 \\ z_2\overline{z}_1 & z_2\overline{z}_2 & z_2\overline{z}_3 & z_2\overline{z}_4 \\ z_3\overline{z}_1 & z_3\overline{z}_2 & z_3\overline{z}_3 & z_3\overline{z}_4 \\ z_4\overline{z}_1 & z_4\overline{z}_2 & z_4\overline{z}_3 & z_4\overline{z}_4\end{array}\right)
\end{displaymath} 
where the nonzero $\lambda = (z_1,z_2,z_3,z_4) \in \mathbb{C}^4$ satisfy
\begin{equation}
||z_1||^2+||z_2||^2+||z_3||^2+||z_4||^2=1.
\end{equation}
It is not difficult to see that $\textrm{det}(p)=0$.  Such conditions on $p$ are equivalent to those discussed by Witten \cite{40} for lightlike momentum vectors $p_{a\dot{a}}$ of the form:
\begin{equation}
p_{a\dot{a}}=\lambda_a\tilde{\lambda}_{\dot{a}}.
\end{equation}
Therefore, given a $\lambda$ that satisfies (2), we acquire a lightlike vector $p$, from which we can determine a polarization vector, up to gauge transformation.  This is ample information to write scattering amplitudes in $\mathbb{CP}^3$ twistor theory \cite{40}.\\
\indent We arrive at a representation of $SL(4,\mathbb{C})$ on $\mathfrak{h}_4(\mathbb{C})$ \cite{6} by first considering the Lie algebra $\mathfrak{sl}(4,\mathbb{C})$, generated by $4\times 4$ traceless complex matrices.  The fundamental representation of $\mathfrak{sl}(4,\mathbb{C})$ as linear operators on $\mathbb{C}^4$ is given by
\begin{equation}
a:x\mapsto ax,\qquad x\in\mathbb{C}^4
\end{equation}
for $a\in\mathfrak{sl}(4,\mathbb{C})$.  Tensoring the fundamental representation with its dual, we recover a representation of $\mathfrak{sl}(4,\mathbb{C})$ on the space of $4\times 4$ matrices $M_4(\mathbb{C})$, given by
\begin{equation}
a:x\mapsto ax + xa^*,\qquad x\in M_4(\mathbb{C})
\end{equation}
for a traceless complex matrix $a$.  The sum $ax + xa^*$ gives a Hermitian matrix when $x$ is Hermitian, thus giving a representation of $\mathfrak{sl}(4,\mathbb{C})$ on $\mathfrak{h}_4(\mathbb{C})$.  Exponentiating, we obtain a representation of the group $SL(4,\mathbb{C})$ on $\mathfrak{h}_4(\mathbb{C})$.
\\\\
\textbf{Example 2.2.2 ($\mathbb{CP}^{3|4}$ Twistor String Theory and $\mathfrak{h}_5(\mathbb{C})$)}\\
\indent In the case of the Calabi-Yau supermanifold $\mathbb{CP}^{3|4}$, the twistor coordinates $Z$ are extended as $\mathcal{Z}=(Z^I,\psi^A)$, for $I=1,2, 3$, $A=1,\dots,4$ where $\psi^A$ are fermionic and of charge one with respect to the $U(1)$ gauge field $B$ \cite{40}.  To recover the structure of $\mathbb{CP}^{3|4}$, we use the Jordan algebra $\mathfrak{h}_5(\mathbb{C})$ which decomposes via isomorphism as:
\begin{displaymath}
\mathfrak{h}_5(\mathbb{C})\cong\mathbb{R}\oplus\mathfrak{h}_4(\mathbb{C})\oplus\mathbb{C}^4
\end{displaymath}
\begin{equation}
\left(\begin{array}{cc}\alpha & \psi \\ \psi^* & a \end{array}\right)\mapsto (\alpha,\psi,a)
\end{equation}
where $\alpha\in\mathfrak{h}_4(\mathbb{C})$, $\psi\in\mathbb{C}^4$, and $a\in\mathbb{R}$.
We recover coordinates of the projective twistor space $\mathbb{CP}^3$ from rank one and two projections of $\mathfrak{h}_4(\mathbb{C})$.  The copy of $\mathbb{C}^4$ provides four-component complex spinors $\psi^A$.  Therefore, using the Jordan algebra $\mathfrak{h}_5(\mathbb{C})$, we see that twistor coordinates for $\mathbb{CP}^{3|4}$ can be embedded in the space of $5\times 5$ complex Hermitian matrices.  This suggests that the B-model \cite{40} is related an $\mathfrak{h}_5(\mathbb{C})$ matrix model.
\subsection{Spin Factors and Minkowski Spacetime} 

Let $V$ be an $n$-dimensional real inner product space $V$.  The \emph{spin factor} $J(V)$ is the Jordan algebra freely generated by $V$ modulo the relation \cite{6}
\begin{equation}
v^2 = ||v||^2.
\end{equation}
$J(V)$ is isomorphic to $V\oplus \mathbb{R}$ via the product
\begin{equation}
(v,\alpha)\circ (w,\beta)=(\alpha w + \beta v,\hspace{.1cm}<v,w> + \hspace{.1cm}\alpha \beta).
\end{equation}
The spin factor $J(V)$ is naturally equipped with a symmetric bilinear form of signature $(n,1)$, the Minkowski metric:
\begin{equation}
(v,\alpha)\cdot (w,\beta)= \hspace{.1cm}<v,w> - \hspace{.1cm}\alpha\beta.
\end{equation}
This allows us to regard $J(V)\cong V\oplus \mathbb{R}$ as Minkowski spacetime, with $V$ as space and $\mathbb{R}$ as time.
The \emph{lightcone} $C(V)$ consists of all nonzero $s\in J(V)$ such that $s\cdot s = 0$.  A \emph{light ray} is a 1-dimensional subspace of $J(V)$ spanned by an element of $C(V)$.  The space of all light rays is called the \emph{heavenly sphere} $S(V)$, which is the projective space built from the Jordan algebra $J(V)$ \cite{6}.\\
\indent To generalize the construction for the other simple formally real Jordan algebras takes some modification.  Extending the construction in \cite{6}, we define the lightcone $C_{\mathfrak{h}(\mathbb{K})}$ to consist of all nonzero $\Phi\in\mathfrak{h}_{m+1}(\mathbb{K})$ such that $\Phi\ast \Phi = 0$.  We define a light ray as a rank one projection in $\mathfrak{h}_{m+1}(\mathbb{K})$, which leaves the heavenly sphere to be defined as the projective space $\mathbb{KP}^m$. 
\section{Octonionic Matrix Model Geometry}
\subsection{The Geometry of the Exceptional Cubic Matrix Model}
The matrix model proposed by Smolin \cite{3}, called the \emph{exceptional cubic matrix model} has degrees of freedom in $\mathfrak{h}_3(\mathbb{O})\times\mathcal{G}$ and is defined by the action:
\begin{equation}
S=\frac{k}{4\pi}f_{ijk}t(X^i,\rho\circ X^j, \rho^2\circ X^k)
\end{equation}
where $f_{ijk}$ are structure constants of $\mathcal{G}$, $t(.,.,.)$ is a trilinear form, and
\begin{equation}
X^{\mu}=\left(\begin{array}{ccc}a_1 & \varphi_1 & \overline{\varphi}_2 \\ \overline{\varphi}_1 & a_2 & \varphi_3 \\ \varphi_2 & \overline{\varphi}_3 & a_3 \end{array}\right) \qquad \qquad a_i \in \mathbb{R} \quad\varphi_j \in \mathbb{O}.
\end{equation}
The $X^{\mu}$ are Hermitian elements of $\mathfrak{h}_3(\mathbb{O})$, the exceptional Jordan algebra.\\
\indent  As the exceptional cubic matrix model is based on Hermitian elements of $\mathfrak{h}_3(\mathbb{O})$, the geometry of the model includes the geometry of $\mathfrak{h}_3(\mathbb{O})$.  As expected, trace one and trace two projections in $\mathfrak{h}_3(\mathbb{O})$ give points and lines of the octonionic projective plane $\mathbb{OP}^2$.  It was shown \cite{11} that any projection with trace one has the form
\begin{displaymath}
p=vv^{\dagger}=\left(\begin{array}{c}\varphi_1 \\ \varphi_2 \\ \varphi_3 \end{array}\right)\left(\begin{array}{ccc}\overline{\varphi}_1 & \overline{\varphi}_2 & \overline{\varphi}_3 \end{array}\right) = \left(\begin{array}{ccc}\varphi_1\overline{\varphi}_1 & \varphi_1\overline{\varphi}_2 & \varphi_1\overline{\varphi}_3 \\ \varphi_2\overline{\varphi}_1 & \varphi_2\overline{\varphi}_2 & \varphi_2\overline{\varphi}_3 \\ \varphi_3\overline{\varphi}_1 & \varphi_3\overline{\varphi}_2 & \varphi_3\overline{\varphi}_3 \end{array}\right)
\end{displaymath}
where the nonzero $v = (\varphi_1,\varphi_2,\varphi_3) \in \mathbb{O}^3$ satisfy
\begin{equation}
(\varphi_1\varphi_2)\varphi_3 = \varphi_1(\varphi_2\varphi_3),\qquad\qquad ||\varphi_1||^2+||\varphi_2||^2+||\varphi_3||^2+||\varphi_4||^2=1.
\end{equation}
Any projection with trace two takes the form $I-p$, where $p$ has trace one \cite{6}.  This gives us a one-to-one correspondence between points and lines in $\mathbb{OP}^2$.  Even more, $\mathbb{OP}^2$ is self-dual \cite{6}.  Lines in $\mathbb{OP}^2$ are copies of $\mathbb{OP}^1$.  For any two distinct points in $\mathbb{OP}^2$, there is a unique $\mathbb{OP}^1$ on which they both lie.  For any two distinct $\mathbb{OP}^1$ lines there is a unique point lying on both of them.\\
\indent We recover trace one projections naturally through the eigenvalue problem for $\mathfrak{h}_3(\mathbb{O})$ \cite{9}.  For elements of $\mathfrak{h}_3(\mathbb{O})$, we find three real eigenvalues when the eigenvalue problem is written as
\begin{equation}
\Phi\circ p = \lambda p
\end{equation}
for $p$ a trace one projection matrix.  The characteristic equation for this problem takes the form
\begin{equation}
-\textrm{det}(\Phi - \lambda I)= \lambda^3 - (\textrm{tr}\Phi)\lambda^2 + \sigma(\Phi)\lambda - (\textrm{det}\Phi)I=0
\end{equation}
where $\sigma(\Phi)=\textrm{tr}(\Phi\ast\Phi)$.  If there are no repeated solutions we acquire the decomposition
\begin{equation}
\Phi=\sum^3_{i=1}\lambda_i p_i
\end{equation}
where $p_i$ are trace one projections of $\mathfrak{h}_3(\mathbb{O})$ that satisfy the condition $p_i\circ p_j=0$.  Quantum mechanically, this gives a well-defined position $(\lambda_1,\lambda_2,\lambda_3)$ and momentum $(p_1,p_2,p_3)$ for a dynamical variable $\mathcal{A}_{\Phi}$ in three-dimensional space.  In the next section, we will interpret such three-dimensional dynamical variables in terms of bound states of D-branes and explain how the associated quantum mechanics produces an exceptional twistor string theory. 
\subsection{Relation to the BFSS Matrix Model}   
\indent To relate the exceptional cubic matrix model to the BFSS matrix model \cite{5,12}, we invoke the isomorphism \cite{6}:
\begin{displaymath}
\mathfrak{h}_3(\mathbb{O})\cong\mathbb{R}\oplus\mathfrak{h}_2(\mathbb{O})\oplus\mathbb{O}^2
\end{displaymath}
\begin{equation}
\left(\begin{array}{cc} X & \theta \\ \tilde{\theta} & a \end{array}\right)\mapsto (X,a,\theta).
\end{equation}
$X\in\mathfrak{h}_2(\mathbb{O})$ and $\theta\in\mathbb{O}^2$ are identified with a vector and spinor in 9+1-dimensional Minkowski spacetime respectively, with an extra real scalar.  This is supported by the fact that the spinor representation of $\mathfrak{so}(9)$ splits as $\mathbf{8}_c\oplus\mathbf{8}_s$ when restricted to $\mathfrak{so}(8)$, giving the nine dimensional spinor isomorphism
\begin{equation}
S_9\cong \mathbb{O}^2.
\end{equation}
The spin factor $\mathfrak{h}_2(\mathbb{O})$ splits via isomorphism as
\begin{equation}
\mathfrak{h}_2(\mathbb{O})\cong (\mathbf{8}_v\oplus \mathbb{R})\oplus \mathbb{R}
\end{equation}
giving a representation of 9+1-dimensional Minkowski spacetime.  The nine spatial coordinates are encoded in the $X^i\in (\mathbf{8}_v\oplus \mathbb{R})$, with their superpartners given by $\theta\in\mathbb{O}^2$. The $\mathfrak{so}(8)$ representations $\mathbf{8}_v$, $\mathbf{8}_s$ and $\mathbf{8}_c$ are mixed through the triality generators of $\mathfrak{h}_3(\mathbb{O})$, which act as:
\begin{equation}
\rho \circ \Phi = \left(\begin{array}{ccc}a_2 & \varphi_3 & \overline{\varphi}_1 \\ \overline{\varphi}_3 & a_3 & \varphi_2 \\ \varphi_1 & \overline{\varphi}_2 & a_1 \end{array}\right)
\end{equation}
\indent As the matrices $X^i$, $i=1,...,9$ are related to spacetime coordinates, it is said \cite{5,6,36} that they are simultaneously diagonalizable only along sectors where $[X^i,X^j]$ vanishes.  The eigenvalues are interpreted as transverse D0-brane positions, with the branes connected by fundamental strings.  In the present Jordan algebraic formalism, there is no need for the commutators $[X^i,X^j]$ to vanish.  This is because the $X^i$ are elements of $\mathfrak{h}_2(\mathbb{O})$, embedded in the exceptional Jordan algebra $\mathfrak{h}_3(\mathbb{O})$, and we have already seen how to recover real eigenvalues for elements $\Phi\in\mathfrak{h}_3(\mathbb{O})$.  In fact, elements of $\mathfrak{h}_3(\mathbb{O})$ can be diagonalized with $F_4$ transformations, as $F_4$ is the automorphism group of the exceptional Jordan algebra $\mathfrak{h}_3(\mathbb{O})$ \cite{9,10}.  The diagonalization of elements of $\mathfrak{h}_2(\mathbb{O})$ is an integral step in this process \cite{9}.  Therefore, of the three real eigenvalues found for elements of $\mathfrak{h}_3(\mathbb{O})$, two correspond to elements of $\mathfrak{h}_2(\mathbb{O})$.  This gives only two dynamical degrees of freedom for the BFSS matrix model, rather than nine.\\
\indent The simultaneous diagonalization of the $X^i$ over the projective space $\mathbb{OP}^2$ gives two eigenvalues with two corresponding rank one projection matrices $p_{\mu}$.  The $p_{\mu}$ are points of $\mathbb{OP}^2$, an octonionic generalization of twistor projective space.  From the last section recall that all rank one projection matrices of $\mathfrak{h}_3(\mathbb{O})$ take the form:
\begin{equation}
p=vv^{\dagger}
\end{equation} 
so can be interpreted as the octonionic analog of Witten's lightlike bi-spinors \cite{40}.  By the axioms of a projective plane, given two points in $\mathbb{OP}^2$ there is a unique $\mathbb{OP}^1$ on which they lie.  Therefore the diagonalization of the $X^i$ gives two position values, two lightlike bi-spinors, and a unique $\mathbb{OP}^1$ eight-sphere on which the bi-spinors lie.  The BFSS matrix theory interpretation is that the two eigenvalues give positions of D0-branes, which as a bound state form a supergraviton \cite{5}.  Through the lens of twistor string theory \cite{40, 41}, it is tempting to regard $\mathbb{OP}^1$ as the octonionic generalization of a holomorphic or algebraic curve, interpreted as the worldsheet of a string.  Whatever the case, the full intepretation must take into account all three real eigenvalues of elements of $\mathfrak{h}_3(\mathbb{O})$, as well as the corresponding points in $\mathbb{OP}^2$.  This would give the physical interpretation of Smolin's exceptional cubic matrix model.\\
\indent In closing, the BFSS matrix model describes dynamics in the projective twistor space $\mathbb{OP}^1$.  The relevant spinors are two-component elements of $\mathbb{O}^2$.  We have shown that both can be embedded in the exceptional Jordan algebra $\mathfrak{h}_3(\mathbb{O})$.  This suggests that Smolin's matrix model describes an octonionic generalization of the topological B-model with target space $\mathbb{OP}^{1|2}$, for $\mathcal{Z}=(Z^I,\psi^A)$, $I=1$, $A=1,2$.  The relevant conformal group is $SL(2,\mathbb{O})$, which has a representation on $\mathfrak{h}_2(\mathbb{O})$ \cite{7}, giving $PSL(2,\mathbb{O})$ acting as conformal transformations of $\mathbb{OP}^1$.  
\subsection{The $E_6$ Matrix Model}
Inspired by Smolin's matrix model, Ohwashi formulated a matrix model based on the exceptional Jordan $C^*$-algebra $\mathfrak{h}_3(\mathbb{C}\otimes\mathbb{O})$.  The Chern-Simons type action is defined as
\begin{equation}
S=f_{ijk}c(\Omega^{[i},\rho\circ \Omega^j, \rho^2\circ \Omega^{k]})
\end{equation}
where \emph{c}(.,.,.) is the cubic form, $\Omega^{\mu}\in\mathfrak{h}_3(\mathbb{C}\otimes\mathbb{O})$, and $f_{ijk}$ are structure constants.  The $\Omega^{\mu}$ are $3\times 3$ matrices over the bioctonions of the form
\begin{equation}
\Omega^{\mu}=\left(\begin{array}{ccc}z_1 & \eta_1 & \eta_2 \\ \tilde{\eta}_1 & z_2 & \eta_3 \\ \tilde{\eta}_2 & \tilde{\eta}_3 & z_3 \end{array}\right) \qquad \qquad z_i \in \mathbb{R} \quad\eta_j \in \mathbb{C}\otimes\mathbb{O},
\end{equation} 
where the $\tilde{\eta_j}$ denotes elements under bioctonionic conjugation \cite{2,22}.\\
\indent  The relevant projective space in this case is the complex Moufang plane, $(\mathbb{C}\otimes\mathbb{O})\mathbb{P}^2$.  Points of $(\mathbb{C}\otimes\mathbb{O})\mathbb{P}^2$ are those $\Omega\in \mathfrak{h}_3(\mathbb{C}\otimes\mathbb{O})$ that satisfy $\Omega\ast \Omega = 0$.  Given $\Omega^1,\Omega^2\in(\mathbb{C}\otimes\mathbb{O})\mathbb{P}^2$, the distance between them or the transition probability $\Pi_{1,2}$ is given by \cite{17,22}
\begin{equation}
\Pi_{1,2}=tr(\Omega^1\circ \Omega^2).
\end{equation}  
The distance is invariant under $E_6$, the isometry group of the complex Moufang plane \cite{7}.\\
\indent If we regard $(\mathbb{C}\otimes\mathbb{O})\mathbb{P}^2$ as an exceptional twistor space, it contains all the lightlike elements of the $C^*$-algebra $\mathfrak{h}_3(\mathbb{C}\otimes\mathbb{O})$. $\mathfrak{h}_3(\mathbb{C}\otimes\mathbb{O})$ is a commutative, nonassociative algebra, so describes a nonassociative geometry that is a bioctonionic extension of Minkowski space.  This geometry includes all the geometry of the matrix models described in the last section, being that $\mathfrak{h}_3(\mathbb{O})$ is the self-adjoint part of $\mathfrak{h}_3(\mathbb{C}\otimes\mathbb{O})$.\\
\indent  It is known \cite{9} that $\mathfrak{h}_3(\mathbb{C}\otimes\mathbb{O})$ is the 27-dimensional representation of the exceptional group $E_6$.  The relevant spinors in this case are two-component elements of $(\mathbb{C}\otimes\mathbb{O})^2$.  These furnish a bioctonionic representation of 32-dimensional Majorana spinors \cite{4}.  The Grassmann algebra properties of these bioctonionic spinors was studied in \cite{2}.  In the $E_6$ matrix model, these spinors are mixed with bosonic degrees of freedom through cycle mappings or triality generators \cite{4}.  It is likely the $E_6$ matrix model describes a topological model with target space $(\mathbb{C}\otimes\mathbb{O})\mathbb{P}^{1|2}$ for $\mathcal{Z}=(Z^J,\Psi^B)$, $J=1$, $B=1,2$.  The study of such a generalized twistor string theory is beyond the scope of this paper, but would surely yield interesting results.\\
\section{Conclusion}
In this paper we have shown that Jordan algebras provide a natural means for the study of twistor theory in a matrix formalism.  Through this formalism, the geometries of the exceptional cubic and $E_6$ matrix models were shown to yield exceptional versions of projective twistor space.  This generalizes twistor string theory in such a way that \textit{twistor matrix theory} seems a more proper title.  Future studies of twistor matrix theory will reveal if the formalism is suitable for the study of scattering processes.  In the $F_4$ case, it is likely scattering amplitudes can be written in terms of rank one projections of $\mathfrak{h}_3(\mathbb{O})$.  In the $E_6$ case, however, the construction of rank one idempotents from spinors remains an open problem.

\end{document}